\documentclass[a4paper]{ESASPCS13Style}
\usepackage{epsfig}

\begin{document}

\title{ WAVELENGTH SHIFTS IN SOLAR-TYPE SPECTRA }
\author{ D.Dravins, L.Lindegren, H.-G.Ludwig, \and S.Madsen}
 \institute{ Lund Observatory, Box 43, SE-22100 Lund, Sweden }
\maketitle 

\begin{abstract}

Spectral-line displacements away from the wavelengths naively expected from the Doppler shift caused by stellar radial motion may originate as convective shifts (correlated velocity and brightness patterns in the photosphere), as gravitational redshifts, or perhaps be induced by wave motions.  {\it Absolute lineshifts}, in the past studied only for the \object{Sun}, are now accessible also for other stars thanks to astrometric determination of stellar radial motion, and spectrometers with accurate wavelength calibration.

Comparisons between spectroscopic apparent radial velocities and astrometrically determined radial motions reveal greater spectral blueshifts in F-type stars than in the Sun (as theoretically expected from their more vigorous convection), further increasing in A-type stars (possibly due to atmospheric shockwaves).

Work is in progress to survey the spectra of the Sun and several solar--type stars for "unblended" photospheric lines of most atomic species with accurate laboratory wavelengths available.  One aim is to understand the ultimate information content of stellar spectra, and in what detail it will be feasible to verify models of stellar atmospheric hydrodynamics.  These may predict bisectors and shifts for widely different classes of lines, but there will not result any comparison with observations if such lines do not exist in real spectra, or are too blended for meaningful measurement.

An important near--future development to enable a further analysis of stellar surface structure will be the study of wavelength variations across {\it spatially resolved stellar disks}, e.g., the center--to--limb wavelength changes along a stellar diameter, and their spatially resolved time variability.

\end{abstract}

\section{ Really understanding stellar atmospheres}

Improved understanding of physical processes in stellar atmospheres comes from the interplay between theory and observations.  The validity of, e.g., hydrodynamic simulations of stellar photospheres can be tested by whether they succeed in reproducing observed shapes and asymmetries of photospheric spectral lines.  On one hand, the success of such models (at least for solar-type stars) is encouraging in suggesting that many physical processes probably are beginning to get well understood.  On the other hand, however, similar spectral--line shapes may sometimes be predicted by models with different parameter combinations, showing the need for further observational constraints.  New diagnostic tools are needed to, e.g., segregate 2-dimensional models from 3-dimensional ones, to clarify effects of non-LTE, or of 3-dimensional radiative--transfer effects in line formation.

The need for additional diagnostic tools is made more urgent whenever significant predictions are obtained from hydrodynamic models which disagree with, e.g., simple one--dimensional ones.  For example, any hydrodynamic model that is predicting specific chemical abundances (e.g., Steffen \& Holweger 2003, Asplund et al.\ 2004), should be checked whether it does precisely reproduce observed line parameters also for various other species. 

This paper discusses absolute wavelength shifts of photospheric spectral lines.  This denotes the wavelength displacements away from the positions the lines naively would be expected to have, had they been affected only by the Doppler shift caused by the relative star--Earth motion.  In the past, such absolute lineshifts could be studied only for the Sun (since the relative Sun--Earth motion, and the ensuing Doppler shift is known from planetary--system dynamics, and does not rely on measurements in the solar spectrum).  For other stars, this has now become possible thanks to three separate developments: (a) Accuracies reached in space astrometry (especially the Hipparcos mission) have made it possible to determine stellar radial motion from second--order effects in astrometry only, without using any spectroscopy; (b) The availability of high--resolution spectrometers with accurate wavelength calibration (such as designed for exoplanet searches), and (c) Accurate laboratory wavelengths for several atomic species.  Already the two latter points (b) and (c) permit studies of {\it differential} shifts between different classes of lines in one and the same star even if, in the absence of astrometric radial velocities, the shifts can not be placed on an absolute scale.

\begin{figure*}[ht]
  \begin{center}
    \epsfig{file= 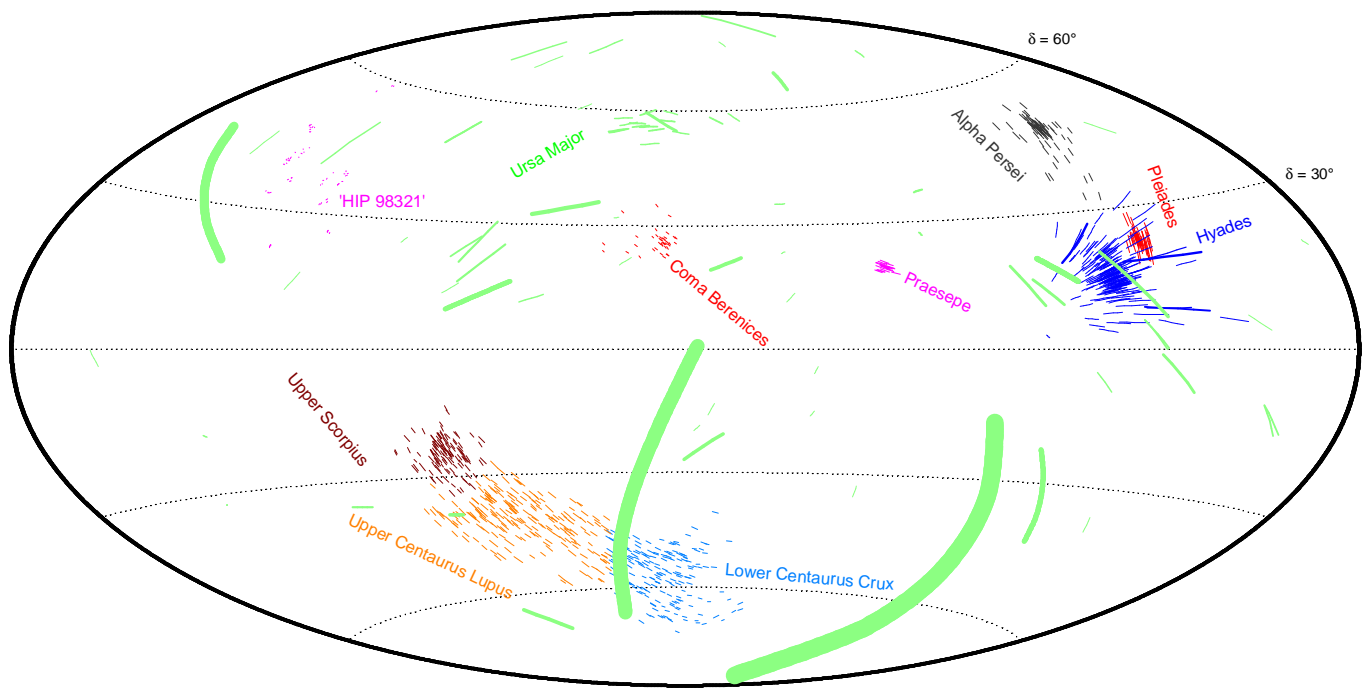, width=18cm}
  \end{center}
\caption{Nearby open clusters, for whose stars astrometric radial--velocity solutions have been obtained from Hipparcos data.  Proper motions are shown extrapolated for 200,000 years. Best radial--velocity accuracy is obtained in rich nearby clusters with large angular extent, and large proper motions. However, the accuracy in the largest associations (Ursa Major, Scorpius--Centaurus) is limited by the partly unknown expansion of these systems. Stellar paths in the Ursa Major group (shaded) cover large areas of the sky. The thickness of the proper--motion vectors is inversely proportional to stellar distance: the closest star is Sirius and the two next ones are faint red dwarfs. Proper motions vary greatly among different clusters. (Madsen et al.\ 2002) \label{fig1}}
\end{figure*}

Asymmetries and wavelength shifts of photospheric absorption lines in particular originate from correlated velocity and brightness patterns (stellar granulation): rising (blueshifted) elements are hot (bright), and convective blueshifts normally result from a larger contribution of such blueshifted photons than of redshifted ones from the sinking and cooler (darker) gas.  The exact amount of shift differs among lines with different conditions of formation.  For example, high--excitation lines may form predominantly in the hottest (also the most rapidly rising, and most blueshifted) elements, and thus show a more pronounced blueshift.  By contrast, lines formed in high-lying layers of convective overshoot may experience an inverted correlation, instead resulting in a convective redshift.  For overviews of processes causing such shifts, see Dravins (1999), Asplund et al.\ (2000), Allende Prieto et al.\ (2002), and references therein.

\section{Radial velocities without spectroscopy!}

Traditionally, stellar radial velocities have of course been determined by spectroscopy, utilizing the Doppler effect.  The advent of high--accuracy (milli--arcsecond, or better) astrometric measurements now permits radial velocities to be obtained also by purely geometric methods.  Such {\it astrometric radial velocities}, directly giving the motion of the stellar center--of--mass, are independent of phenomena affecting stellar spectra, such as stellar pulsation, convection, rotation, winds, isotopic composition, pressure, and gravitational potential.  Among the several astrometric techniques (Dravins et al.\ 1999b), one method reaches sub-km~s$^{-1}$ accuracies already with existing data.  This "moving--cluster method" is based on the circumstance that stars in suitable open clusters move through space with a common [average] velocity vector.  The radial--velocity component makes the cluster appear to contract or expand due to its changing distance.  This relative rate of apparent contraction (observed through stellar proper motions) equals the relative rate of change in distance, which can be converted to a linear velocity (in km~s$^{-1}$) if stellar distances are known from trigonometric parallaxes.  Once the cluster's space velocity is known, the radial velocity for any member star follows by projecting the cluster's velocity vector onto the line of sight.  In a sense, the method can be regarded as an inversion of the classical moving--cluster problem, where the distance is derived from proper motions and [spectroscopic] radial velocities: here, by instead first knowing the distances, the radial velocities follow.

\section{Lineshifts intrinsic to stellar atmospheres}

On Hipparcos, an observing program was carried out for stars in moving clusters, yielding astrometric radial--velocity solutions for more than 1,000 stars (Madsen et al.\ 2002), including about one hundred stars in the nearby Hyades and Ursa Major groups (Figure 1).  The error budgets are somewhat complex, but accuracies may reach around 0.3 km~s$^{-1}$ (Dravins et al.\ 1999b, Madsen 2003).

The differences between astrometrically determined radial velocities (giving motions of the stellar centers of mass) and the apparent spectroscopic velocities of different features reveal lineshifts intrinsic to stellar atmospheres, such as convective and gravitational lineshifts.  Figure 2 for Hyades stars identifies three regions of difference between astrometric and spectroscopic velocities.

\begin{figure}[ht]
  \begin{center}
    \epsfig{file= 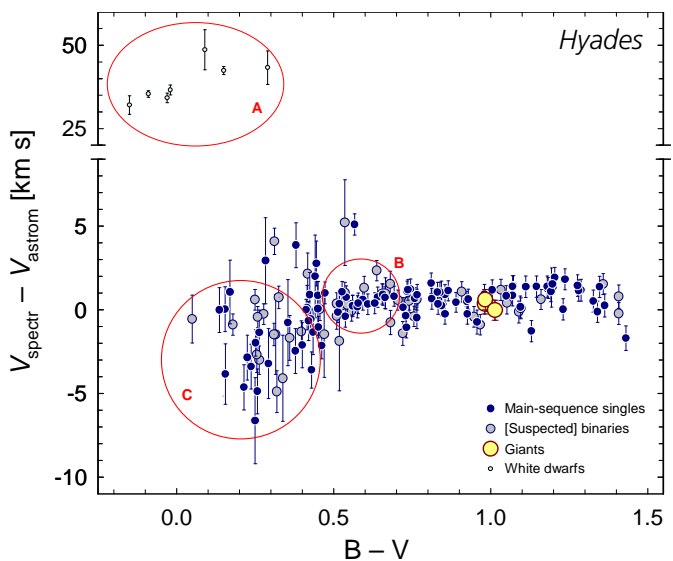, width=8.8cm}
  \end{center}
\caption{ The Hyades: Differences between spectroscopic radial--velocity values from the literature, and astrometric determinations.  Gravitational redshifts of white--dwarf spectra place them far off main--sequence stars (A).  An increased blueshift of spectral lines in stars somewhat hotter than the Sun (B--V $\simeq 0.4-0.7$) is theoretically expected due to their more vigorous surface convection, causing greater convective blueshifts (B).  The spectra of still earlier--type stars appear systematically blueshifted (C).  The error bars show the combined spectroscopic and astrometric errors, of which the latter are negligible by comparison.  (Adapted from Madsen et al.\ 2002) \label{fig2}}
\end{figure}

\subsection{Gravitational redshifts}

\begin{figure}[ht]
  \begin{center}
    \epsfig{file= 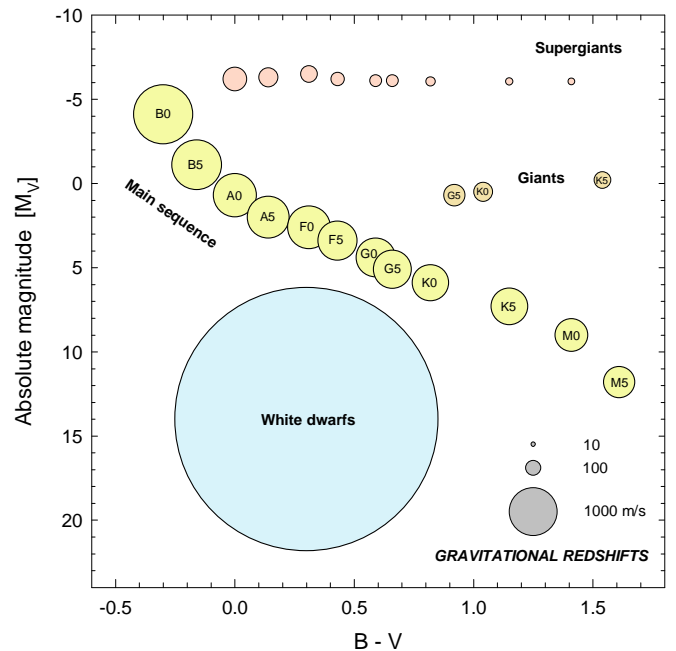, width=8.7cm}
  \end{center}
\caption{ Gravitational redshifts throughout the Hertzsprung--Russell diagram are expected to change by three orders of magnitude between white dwarfs ($\simeq$ 30 km~s$^{-1}$) and supergiants ($\simeq$ 30 m~s$^{-1}$).  Symbol area is proportional to the amount of shift (Dravins et al.\ 1999a) \label{fig3}}
\end{figure}

Region (A) in Figure 2 identifies stellar spectra that are redshifted by $\simeq$ 30~km~s $^{-1}$ relative to the "expected" Doppler shift, revealing the pronounced gravitational redshift in the Hyades white dwarfs.  The gravitational redshift, $V_{\rm grav} =GM/Rc$, involves the constant of gravitation, the speed of light, and the stellar mass and radius.

Using accepted solar values, the value is 636 m~s$^{-1}$ for light escaping from the solar photosphere to infinity, and 633 m~s$^{-1}$ for light intercepted at the Earth.  A spectral line formed at chromospheric heights (30 Mm, say), will have this shift decreased by some 20 m~s$^{-1}$, and a coronal line by perhaps 100 m~s$^{-1}$.  For other stars, the shift scales as ($M/M_{\odot }$)($R_{\odot }/R$).

It is believed that this shift can be reasonably well predicted from stellar structure models (for well--studied single main--sequence stars to an accuracy of perhaps 50 m~s$^{-1}$).  Figure 3 shows expected gravitational redshifts predicted from standard models.  The shift does not vary much on the main sequence between A5 V and K0 V ($\simeq$~650 m~s$^{-1}$), but reaches twice that value for more massive early-B stars with $M \simeq$~10$~M_{\odot }$.  The convective lineshift, $V_{\rm conv}$, is expected to be about $-$1000 m~s$^{-1}$ for F5~V, $-$400 m~s$^{-1}$ for the Sun, and $-$200 m~s$^{-1}$ for K0~V (Dravins 1999, and references therein).

\subsection{ Modeling of convective lineshifts }

\begin{figure}[ht]
  \begin{center}
    \epsfig{file= 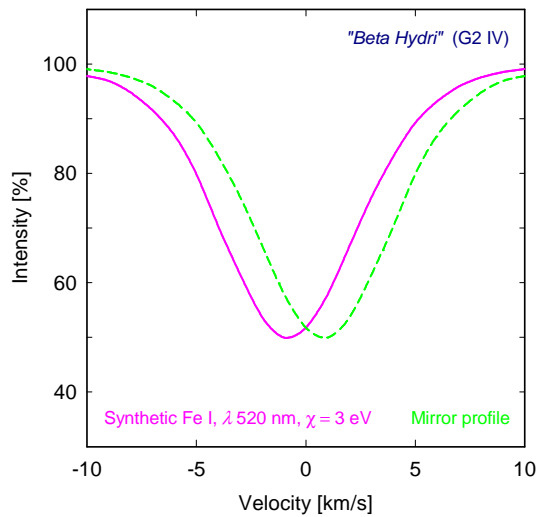, width=7cm}
  \end{center}
\caption{ A model with solar temperature but one quarter of its surface gravity (corresponding to the subgiant \object{$\beta$~Hyi}, G2~IV), has smaller photospheric pressure but greater granular velocities (by a factor of 1.5--2), since the same energy flux as in the Sun must be carried by lower-density gas.  Its resulting convective blueshift is somewhat greater than in the Sun, here illustrated by a synthetic \ion{Fe}{I} $\lambda$~520 nm line, mirrored about the zero-point wavelength (adapted from Dravins et al.\ 1993)
\label{fig4}}
\end{figure}

\begin{figure}
  \begin{center}
    \epsfig{file= 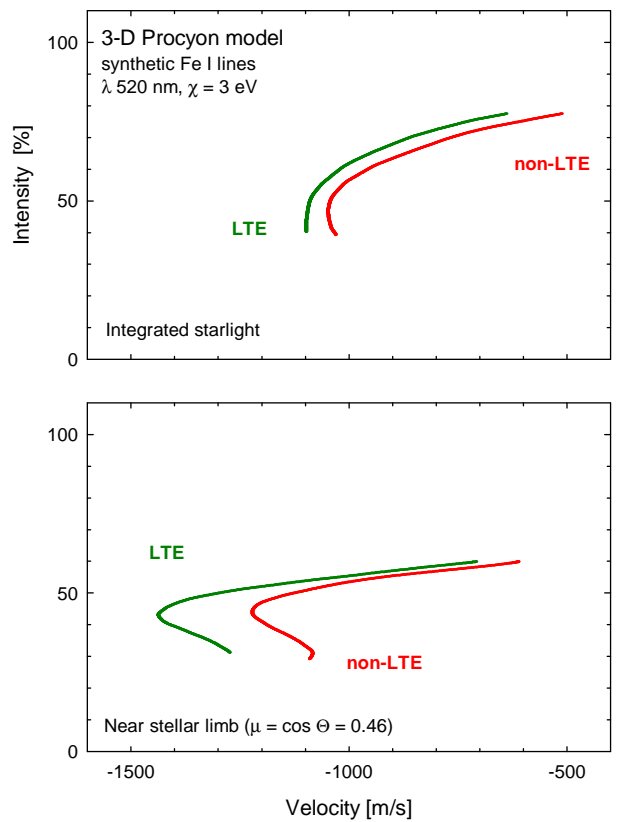, width=8cm}
  \end{center}
\caption{ Non-LTE effects in convective lineshifts, seen in bisectors of \ion{Fe}{I} lines from a 3-D simulation of \object{Procyon}.  Although line shapes and asymmetries are quite similar, there are differences in their shifts.  The smaller shift in non-LTE is caused by a selective reduction of the most blueshifted line components above the hottest granules, whose intense ultraviolet flux ionizes the gas above, thus decreasing contributions from neutral species (adapted from material in Dravins \& Nordlund 1990a) \label{fig5}}
\end{figure}

Synthetic line profiles computed from three--dimensional hydrodynamical models do well reproduce photospheric line shapes and shifts in solar--type stars (e.g., Asplund et al.\ 2000, Allende Prieto et al.\ 2002).  Figure 4 serves to (a) illustrate the magnitude of the effect but (b) also to show that the intensity profile alone (at least on first sight) does not reveal any striking signature of the complex convective velocity fields shaping it.  When going along the main sequence from the Sun towards somewhat earlier--type stars, hydrodynamic models predict more vigorous granulation and enhanced convective blueshifts (e.g., Figure 8 below), apparently consistent with the trend suggested in region (B) of Figure 2.

Three--dimensional hydrodynamic model photospheres are capable of predicting various subtle but potentially observable effects in resulting line profiles.  For example, lineshifts offer another non-LTE diagnostic for synthetic line profiles, since such effects may show up more clearly in shifts than in either profiles or asymmetries.  Although line {\it shapes} may well be similar in LTE, line {\it shifts} can be different (Figure 5).  For lines in the visual from neutral metals in solar--type stars, non-LTE calculations normally produce smaller convective blueshifts because the fastest upflows occur in the hottest granular elements.  These emit strong ultraviolet flux, ionizing the gas above, thus decreasing the most blueshifted line absorption contributions from neutral species.  Close to the stellar limb, the non-LTE effects may become more significant, an effect of "corrugated" stellar surfaces, further discussed in Section 5 below.

\section{ Solar lineshift patterns}

\begin{figure}[ht]
  \begin{center}
    \epsfig{file= 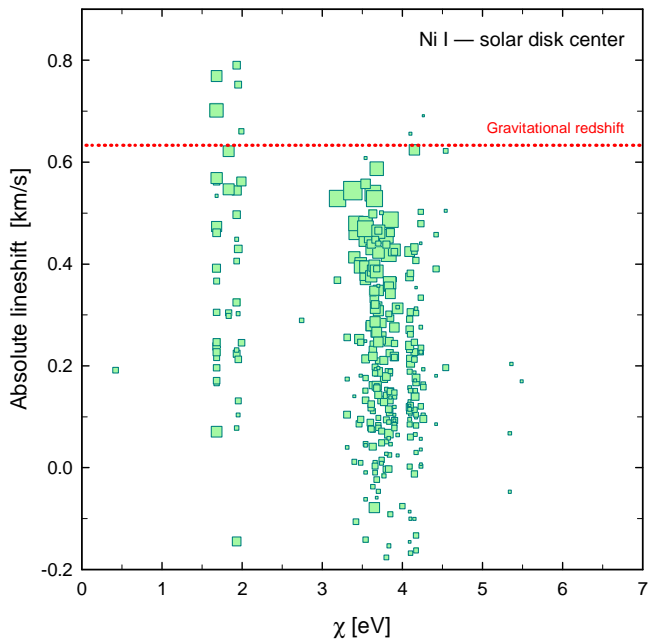, width=8.5cm}
  \end{center}
\caption{ Solar lineshift patterns for \ion{Ni}{I} lines in the visual.  The data are reduced for the Doppler shift due to the relative Sun--Earth motion, and the clustering of lines blueward from the solar gravitational redshift is a signature of granular convection.  Symbol size reflects laboratory intensity.  Solar wavelengths: Allende Prieto \& Garc{\'\i}a L\'{o}pez 1998; Laboratory data: Litz\'{e}n et al.\ 1993
\label{fig6}}
\end{figure}

\begin{figure}[ht]
  \begin{center}
    \epsfig{file= 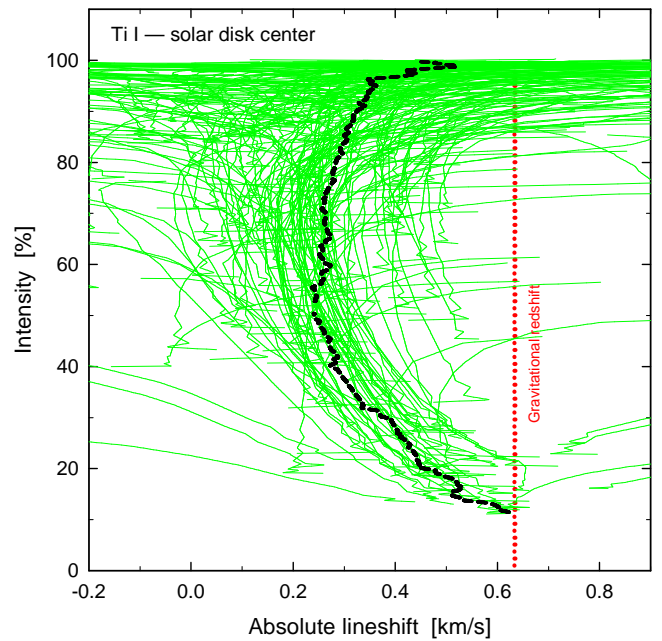, width=8.5cm}
  \end{center}
\caption{Individual bisectors for some 300 relatively unblended \ion{Ti}{I} lines in the visual solar disk--center spectrum, on an absolute wavelength scale.  Despite significant scatter among individual lines, their average (thick dashed line) well defines the convective asymmetry and blueshift relative to the "naively expected" wavelength equal to the solar gravitational redshift. (Dravins, in preparation)
\label{fig7}}
\end{figure}

For the Sun, absolute lineshift studies have been possible since long ago.  Of course, models for the solar atmosphere are developed to a much higher sophistication than those for other stars, implying that more detailed model predictions can be made and tested.  One observational limit is set by systematic errors in wavelength scales which, even in the best solar spectrum atlases, may not be better than perhaps 100 m~s$^{-1}$, comparable to the noise in laboratory wavelengths for low--excitation lines from neutral metals.  The most sensitive tests of various models might be with high--excitation lines, lines from ionized species, or from molecules.  Unfortunately, such lines also appear to be those most sensitive to systematic errors in the determination of laboratory wavelengths.

Traditionally, most analyses of line asymmetries and shifts have been made with  \ion{Fe}{I} lines, exploiting their rich occurrence, as well as the relatively good laboratory data available.  However, the increasing availability of accurate laboratory wavelengths now permits studies for several additional species, including \ion{Ca}{I}, \ion{Co}{I}, \ion{Cr}{I}, \ion{Fe}{II}, \ion{Ti}{I}, and \ion{Ti}{II}.  Not only does this increase the number of accessible lines, but in particular it permits the testing of hydrodynamic models over parameter domains (e.g., excitation and ionization potentials) not reachable in \ion{Fe}{I} only. As an example, Figure 6 shows the solar lineshift pattern for lines from neutral nickel.

Since stellar lines normally are asymmetric, the lineshift is different in different portions of any absorption line.  To determine this lineshift as a function of absorption--depth in the line (usually plotted as the line bisector), is observationally more demanding than a mere determination of the [average] wavelength.  However, also this is becoming practical for non-iron species; an example for \ion{Ti}{I} bisectors at solar disk center is in Figure 7.  Each of the about 300 thin curves is the bisector for one individual \ion{Ti}{I} line in the visual solar disk-center spectrum, placed on an absolute wavelength scale.  This solar spectrum was recorded by Brault et al.\ with the Kitt Peak Fourier-transform spectrometer (see the National Solar Observatory digital library, http://diglib.nso.edu), here with intensities as normalized by Neckel (1999).  Laboratory wavelengths are from Forsberg (1991) and Litz\'{e}n (private comm.).  Even among these relatively unblended lines, there is a noteworthy scatter (due to weak blends, noise in laboratory wavelengths, etc.).  Nevertheless, the average bisector over all these lines does well define the convectively induced asymmetry and blueshift relative to the "naively expected" wavelength equal to the solar gravitational redshift.

\subsection{ Hotter and rapidly rotating stars }

Region (C) in Figure 2 marks hotter (mostly A-type) stars whose spectra appear to be systematically blueshifted by a few km~s$^{-1}$ relative to the "naively" expected value.  Since these earlier--type stars generally are rapidly rotating, it is awkward to identify whether the dependence is primarily on temperature, on rotation, or perhaps also on some other parameter.

In trying to find some explanation for the apparent dependence on stellar rotation, one might consider the increased centrifugal force in rapid rotators which lowers the effective surface gravity.  Analogous to modeled differences between hydrodynamic atmospheric models for dwarfs and lower--gravity subgiants, more vigorous granulation might then be expected to develop (since the convective energy flux must then be carried by a lower-density gas).  The greater velocity amplitudes and temperature contrasts would then become visible in integrated starlight as somewhat enhanced convective blueshifts.

However, a more likely explanation appears to be that these lineshifts are caused by shock--wave propagation.  Although full 3-dimensional hydrodynamic models with ensuing spectral--line synthesis have not yet been developed for A-type stars, various 2-dimensional atmospheric models are becoming available.  One sequence of such, at temperatures near the onset of convection, are being developed by Holweger (2003).  Models for main--sequence stars of 8200 and 9000 K effective temperatures show that the atmosphere is highly dynamic. It moves up and down in a pulsation--like manner and is traversed by numerous shocks which pass up through the photosphere at supersonic velocities of typically 20--30~km~s$^{-1}$.  Rising shocks (moving towards the observer) appear blueshifted; they are accompanied by density enhancements and plausibly must contribute blueshifted spectral--line components.  Given their great amplitudes, it does not seem unreasonable that a statistical bias of the blueshifted profiles could well amount to several km~s$^{-1}$, the amounts indicated in Figure 2.  For a further discussion, see Madsen et al.\ (2003).

For rapidly rotating stars, a tantalizing possibility is the potential of observing {\it meridional flows} across stellar surfaces.  A circulation pattern between the equator and poles will involve rising (or sinking) gases at the equator, with the opposite flows near the poles.  Stars seen equator--on (with a large observed value of $V\sin i$) will then have their spectra influenced by Doppler shifts from the systematic upflow (or downflow) patterns there, while stars seen pole--on (with a small value of $V\sin i$) will show lineshift signatures of the opposite sign.

\section{ Spatially resolved stellar disks}

A significant new development expected already in the near future, is the study of spectral--line variations across {\it spatially resolved stellar disks}, e.g.\ the center--to--limb changes along the equatorial and polar diameters, and their spatially resolved time variability.  The availability of adaptive optics on very large telescopes (and future extremely large ones), long--baseline optical interferometry, and optical aperture synthesis, have the potential to fundamentally change stellar astrophysics, since very many stars will no longer remain unresolved point sources, but finally be revealed as true surface objects.  The combination of such techniques with high--resolution spectroscopy is likely to disclose a rich diversity in spectral line properties, opening up the fine structure of stellar atmospheres to detailed study.

Hydrodynamic models already now give a preview of what might get observed.  The left panels in Figure 8 show bisectors for a strong Fe I line, and those at right for a weak one, both with lower excitation potential $\chi $= 3 eV. The horizontal wavelength scale is absolute (but leaving out gravitational redshifts and effects of stellar rotation).  In the two lower rows, the vertical intensity scale is in units of that at stellar disk center; the somewhat varying intensity levels for the lines near the stellar limb reflect different limb darkening in different stars.  In order of decreasing lineshift, the four stellar models correspond to \object{Procyon} (F5 IV--V); \object{$\beta $~Hydri} (G2 IV); \object{$\alpha $~Cen A} (slightly evolved G2 V), and \object{$\alpha $~Cen B} (K1 V). The bisectors for the Sun are similar in shape but slightly less shifted than those of $\alpha $~Cen A. These data were adopted from models in Nordlund \& Dravins (1990) and Dravins \& Nordlund (1990a; 1990b).

\begin{figure}[ht]
  \begin{center}
    \epsfig{file= 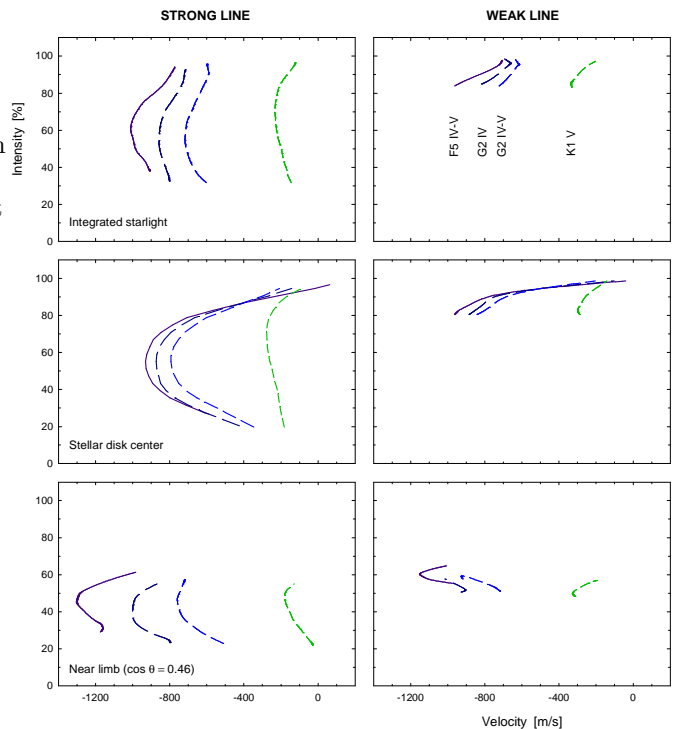, width=8.8cm}
  \end{center}
\caption{ Bisectors for photospheric spectral lines across disks of different solar--type stars, computed from hydrodynamic models of stellar granulation.  Lineshifts differ between different spectral types, and the change in shift across a stellar disk may differ in both size and sign, depending on the detailed stellar surface structure. (Dravins 1999)
\label{fig8}}
\end{figure}

Figure 8 shows that not only may convective blueshifts differ substantially between different spectral types (asymmetries may be quite similar, while shifts differ by factors of several) but, in particular, that the change in shift between stellar disk center toward the limb may {\it{differ in both size and sign}} between different spectral types.  For the Sun, such center--to--limb changes have been known as the "limb effect" since a long time ago, although its origin in terms of convective motions was understood only more recently.

In stars with "smooth" photospheric surfaces (in the optical--depth sense), one may expect the convective blueshift to decrease near the limb, since the vertical convective velocities are then perpendicular to the line of sight, and the horizontal velocities which contribute Doppler shifts appear symmetric. However, stars with "corrugated" surfaces, i.e.\ with "hills" and "valleys" should show an increased blueshift toward the limb.  Although the velocities on the star are horizontally symmetric, one will predominantly see the horizontal velocities on the slopes of those "hills" that are facing the observer. These velocities are approaching the observer and thus appear blueshifted.  The equivalent redshifted components remain invisible behind these "hills", and an enhanced blueshift results.  The effect is somewhat analogous to the asymmetric lines observed from the expanding envelopes of P~Cygni stars.  In Figure 8, this effect is well seen for the F--star model.

Further effects will appear in the {\it{time variability}} of lineshifts across spatially resolved stellar disks: On a "smooth" star, the temporal fluctuations are caused by the random evolution of granular features, all of which are reached by the observer's line of sight.  On a "corrugated" star, observed near its limb, there enters another element of variability in the sense that the changing "corrugations" of the swaying stellar surface sometimes hide some granules from direct view. The result is an enhanced temporal variability of line wavelengths and bisectors (already observed on the Sun), which constitutes another observational parameter for further constraining the hydrodynamics of stellar atmospheres.

\section{ New diagnostics for stellar atmospheres}

A combination of observational, experimental, and theoretical advances now permits absolute lineshifts to be introduced as a diagnostic tool for stellar atmospheres, beyond the previously established ones of line--strength, --width, --shape, and --asymmetry.

On the theoretical side, the reliable prediction of lineshift patterns across the whole Hertzsprung--Russell diagram remains a challenge.  Not only do these require a credible modeling of the atmospheric hydrodynamics, but the lineshifts may vary greatly among different classes of spectral lines (ionized species, molecules, different spectral regions).  This complex dependence actually might be an advantage: Since the exact line displacements depend on so many parameters, stellar atmospheric properties can be constrained by many different wavelength measures for various lines in the same star, or between different stars.

Finally, one should not forget that progress in astronomy (and natural science in general) depends upon the {\it falsification} of theoretical concepts and hypotheses.  If the ambition is to improve a model, the aim is {\it not} to have it fit the observations, but rather to identify situations where the model and observations are in {\it disagreement} (in the case of agreement, there is really not much new that can be learned).  A route for progress in stellar--atmospheres research is then to identify new measures that both can be accurately observed and be precisely predicted by models.  A confrontation between these observational and theoretical quantities may then segregate among competing concepts and guide models toward greater realism and sophistication.

For a further discussion, see links from \\
http://www.astro.lu.se/\~{}dainis

\begin{acknowledgements}

This project is supported by the Swedish Research Council, the Swedish National Space Board, and The Royal Physiographic Society in Lund.

\end{acknowledgements}

\end{document}